\documentclass[extra,mreferee]{gji} 

\usepackage{timet,color}
\usepackage[left]{lineno}
\usepackage{graphicx}
\usepackage[urlcolor=blue,citecolor=black,linkcolor=black]{hyperref}

\title[Microseismicity Detection with DAS]
  {A Semblance-based Microseismic Event Detector for DAS Data}

\author[J. Porras et al.]
  {Juan Porras$^{1,2}$, Davide Pecci$^{1,3}$, Gian Maria Bocchini$^4$, Sonja Gaviano$^1$, \and Michele De Solda$^1$, Katinka Tuinstra$^5$, Federica Lanza$^5$, Andrea Tognarelli$^1$, \and Eusebio Stucchi$^1$ and Francesco Grigoli$^1$ \\
  $^1$ Department of Earth Sciences, University of Pisa, \emph{56124}, Pisa, Italy \\
  $^2$ Department of Earth Sciences, University of Geneva, \emph{1205}, Geneva, Switzerland. Email: juan.porrasloria@unige.ch \\
  $^3$ DESTeC, University of Pisa, \emph{56124}, Pisa, Italy \\
  $^4$ Institute of Geology, Mineralogy and Geophysics, Ruhr University of Bochum, \emph{44801}, Bochum, Germany \\
  $^5$ ETH-Zurich, Swiss Seismological Service (SED), \emph{CH8092}, Zurich, Switzerland
  }
\date{Received 1998 December 18; in original form 1998 November 22}
\pagerange{\pageref{firstpage}--\pageref{lastpage}}
\volume{200}
\pubyear{1998}

\begin{document}

\label{firstpage}

\maketitle

\begin{summary}
Distributed Acoustic Sensing (DAS) is becoming increasingly popular in microseismic monitoring operations. This data acquisition technology converts fiber-optic cables into dense arrays of seismic sensors that can sample the seismic wavefield produced by active or passive sources with a high spatial density, over distances ranging from a few hundred meters to tens of kilometers. However, standard microseismic data analysis procedures have several limitations when dealing with the high spatial (inter-sensor spacing up to sub-meter scale) sampling rates of DAS systems. Here we propose a semblance-based seismic event detection method that fully exploits the high spatial sampling of the DAS data. The detector identifies seismic events by computing waveform coherence of the seismic wavefield along geometrical hyperbolic trajectories for different curvatures and positions of the vertex, which are completely independent from external information (i.e. velocity models). The method detects a seismic event when the coherence values overcome a given threshold and satisfies our clustering criteria. We first validate our method on synthetic data and then apply it to real data from the FORGE geothermal experiment in Utah, USA. Our method detects about two times the number of events obtained with a standard method when applied to 24h of data.   
\end{summary}


\section{Introduction}

Distributed Acoustic Sensing (DAS) is a fiber optic-based data acquisition technology that is gaining popularity in a broad range of seismological applications. From seismic monitoring in volcanic \citep{jousset2022}, volcano-glacial environments \citep{klaasen2021distributed}, offshore areas \citep{shinohara2022} and aftershock scenarios \citep{li2021rapid}, \citep{zeng2022turning}; to downhole \citep{lellouch2021seismic}, surface applications \citep{lindsey2021fiber}, traffic and railway monitoring \citep{wang2021}. Different industrial applications also benefit from the use of DAS, such as induced seismicity monitoring (\citeauthor{webster2013} \citeyear{webster2013}; \citeauthor{karrenbach2019} \citeyear{karrenbach2019}), CO$_2$ sequestration monitoring \citep{daley2013field}, hydraulic stimulation \citep{karrenbach2017hydraulic} and geothermal sites (\citeauthor{lellouch2020comparison} \citeyear{lellouch2020comparison}; \citeauthor{lellouch2021low} \citeyear{lellouch2021low}).

A DAS system operates by sending a laser pulse along a fiber-optic cable. Due to small fluctuations (i.e. inhomogeneities) of the refractive index within the fiber, a portion of the light pulse travelling through the cable is back-scattered. The reflected pulse is sent back along the fiber to the source device that also works as a receiver. This device is commonly known as Interrogator Unit (IU). The IU exploits the Rayleigh back-scattering principle, where the transmitted and the back-scattered pulse have the same frequency, and the phase difference between the two can be translated to dynamic strain along the fiber. When the fiber is unperturbed (i.e. no static or dynamic deformation is occurring), the arrival phase of the back-scattered light is constant. In the presence of an external perturbation (e.g. an incident seismic wave), the fiber length changes and produces a corresponding phase delay in the back-scattered pulse, which is recorded and converted by the IU into strain or strain rate. More technical details about DAS acquisition systems can be found in \citeauthor{lindsey2020broadband} (\citeyear{lindsey2020broadband}) and \citeauthor{paitz2021empirical} (\citeyear{paitz2021empirical}).

The use of DAS technology is constantly increasing in induced seismicity monitoring with borehole installations \citep{lellouch2021seismic}. In such applications, the typical monitoring setting consists of a geophone chain deployed in a deep borehole, allowing to detect a larger number of seismic events with respect to conventional surface seismic networks (\citeauthor{maxwell12} \citeyear{maxwell12}; \citeauthor{rossi23} \citeyear{rossi23}). Despite their extensive use, borehole arrays of geophones have two main limitations: 1) their operational temperature may strongly affect their installation depth, and 2) the need for a dedicated monitoring well (or observation well) that increases operational costs. These limitations are particularly pernicious in monitoring operations of induced seismicity associated with Enhanced Geothermal Systems (EGS). In fact, the hot dry rocks of an EGS may reach temperatures higher than 300 °C, hence deploying conventional geophones down to the reservoir level is not feasible as the electronics of these sensors is not designed to work in such high temperature conditions \citep{zhidong2019}. Last but not least, conventional seismometer deployments sample the seismic wavefield in a sparse network of point measurements, making these sensors of limited use to analyze the complexity of the seismic wavefield in great detail. Alternatively, the fiber-optic cable can resists to high temperatures \citep{zhidong2019}. In addition, DAS systems sample the seismic wavefield with a very high spatial sampling (up to sub-meter scale) over tens of kilometers \citep{wang2020}, providing a more complete picture of the seismic wavefield. Thus, DAS recordings can potentially enable us to detect features of the seismic wavefields that would go unnoticed with conventional recordings. For example, \citeauthor{lindsey2019illuminating} (\citeyear{lindsey2019illuminating}) show that DAS recording were able to capture/image waves from an on-shore earthquake reflected at previously unknown offshore faults located below the deployed fiber-optic cable.

\begin{figure*}
 \vspace{1.5cm}
 \includegraphics{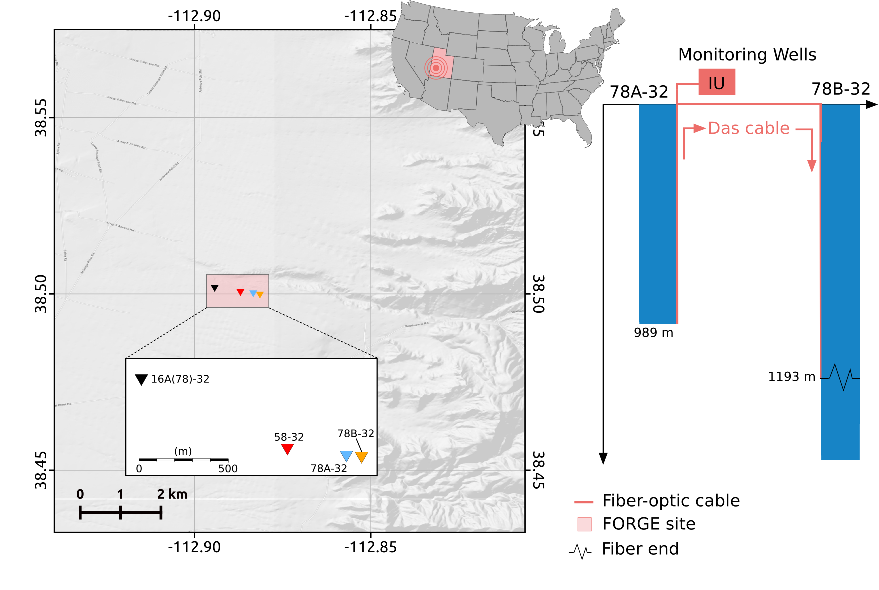}
   \caption{Location map of the FORGE site in Utah, USA. The triangles represent the different wells used for the third stage of the April 2022 stimulation campaign. Well 16A(78)-32 corresponds to the injection well, 58-32 (located at 905 m from the injection well) is a monitoring well with a geophone array used by Geo Energy Suisse to build a seismic catalog. Wells 78A-32 and 78B-32 (located at 1308 m from the injection well) host the fiber-optic of the DAS system. The data used in this study was recorded at well 78B-32. The well-scheme illustrates the design of the DAS system used to monitor the stimulation campaign during April 2022. Wells 78A-32 and 78B-32 are separated by a surface distance of 85 m and connected to allow for simultaneous data acquisition using a single DAS interrogator unit (IU). The system starts recording at the bottom of well 78A-32 and the end of the array is at 1193 m below the surface in well 78B-32.}
   \label{figure4}
\end{figure*}

For these reasons, DAS is a promising technology for the monitoring of induced seismicity associated to EGS operations. Another benefit of this technology is that any production or stimulation well can be turned into an observation well by just installing an optical-fiber behind the casing. This reduces the costs of the monitoring infrastructure but also improves the earthquake detection capability of DAS since the distance from the sensing fiber to the reservoir is reduced. However, a downside of this technology is the massive amount of produced data. An example takes place at the Frontier Observatory for Research in Geothermal Energy (FORGE) site in Utah, USA. This is a dedicated underground field laboratory aiming at developing, testing, and accelerating breakthroughs in EGS technologies to advance the uptake of geothermal resources around the world \citep{oedi_1379} (Fig. \ref{figure4}). At FORGE there have been several DAS acquisition campaigns. For instance, the one in April 2022 \citep{oedi_1379}, which provides the real dataset for this study, used two observation wells, namely 78A-32, and 78B-32. The interrogated sensing optical-fiber was 2482 m long. DAS data were acquired using the following settings: a gauge length of 10 m, a channel spacing of 1 m, and laser launch rate of 20 kHz. Data were saved with a sampling frequency of 4 kHz and in 15 second chunks. Such a configuration produced about 1.3 TB of data per day.

The large datasets produced by DAS systems highlights the problem of data storage. This is particularly true when dealing with weeks or months long seismic monitoring campaigns. A potential solution that could partially solve this challenging problem is to store only the target data, such as waveforms containing only the signal of the seismic events of interest. Unfortunately, standard pick-based seismological techniques do not exploit the high spatial density of DAS data and current seismic event detectors may miss a non negligible number of seismic events. For this reason, using standard detectors during seismic monitoring campaigns with DAS, increases the risk of permanently losing useful information. Another downside of DAS applications is its single-component nature. That is, a DAS system measures only along the axial direction of the fiber and, in borehole geometries, it provides no azimuthal information for earthquake recordings \citep{tuinstra2023locating}.

Since the use of DAS technology for earthquake monitoring is a yet-to-be-established practice, few algorithms for earthquake detection have been developed for DAS data. For instance, \citet{lellouch2019} developed a detection method based on a single-parameter scan of the incidence angle and measurement of data coherence along different possible travel-time curves, given a seismic velocity model. Seismic detection methods based on the measurement of local data coherence have been proposed also for large-N dense networks \citep{li2018high}. Other methodologies of earthquake detection for DAS data based on Convolutional Neural Networks (CNN) have been developed (\citeauthor{binder2019detecting} \citeyear{binder2019detecting}; \citeauthor{stork2020application}\citeyear{stork2020application}, \citeauthor{huot2022detection} \citeyear{huot2022detection} and \citeauthor{zhu2023seismic} \citeyear{zhu2023seismic}). In our opinion, a seismic event detector suitable for DAS data must meet the following criteria: 1) It must exploit the high spatial density of DAS, 2) it must be sensitive but robust at the same time (i.e. low missed and low false event detection rates) and 3) it needs to be computationally fast to be applied in real-time or near real-time. 

In this work we provide a solution to this important problem by implementing HECTOR (coHerence-based Earthquake deteCTOR), a semblance-based seismic event detection method that fully exploits the characteristics of DAS data. We first introduce the theoretical background at the base of our earthquake detector. Next, to evaluate the performance of HECTOR, we test it both on synthetic and real DAS microseismic data acquired during the 2022 EGS stimulation campaign at the Utah FORGE site for which a DAS-based seismic catalogue has been already built by Silixa \citep{silixa}, this is our reference catalogue in this study. Finally, we discuss the performance of the algorithm by comparing our results with the reference catalogue.

\section{Data and Methods}

\subsection{Detection method}

The detection methodology consists of two parts: 1) the first one evaluates the spatial coherence of the seismic waveforms using the semblance function; and 2) the second part performs a clustering analysis of the coherence time series obtained from the first part of the methodology to identify microseismic events. 

\subsubsection{Waveform Coherence}

HECTOR exploits the high spatial sampling of DAS for real-time microseismic monitoring. In particular, for a linear segment of the optical-fiber, HECTOR evaluates the coherence of the seismic waveforms along hyperbolic trajectories by using the Semblance function \citep{neidell1971}. The equation of the hyperbola we use to calculate the coherence is defined as:

\begin{equation}
t_{i}(T,X,C)^{2} = T^{2} + \frac{(x_{i}-X)^{2}}{C^{2}}
\end{equation}

where, $t$ is the time of the trace recorded at the position $x_{i}$ (along the fiber axis), while the coefficients $X$, $T$ and $C$ represent respectively the spatial offset with respect to the vertex of the hyperbola, the time offset, and the curvature. Since we are interested only on positive values, the previous equation can be written as:  

\begin{equation}
t_{i}(T,X,C) = \sqrt{T^{2} + \frac{(x_{i}-X)^{2}}{C^{2}}}
\end{equation}

Finally, the Semblance function $S$ can be written as:

\begin{equation}
S(T,X,C) = \frac{\sum_{j=1}^{N} \left(\sum_{i=1}^{M}A(t_{ij})\right)^2}{M\sum_{j=1}^{N}\sum_{i=1}^{M}A(t_{ij})^2} \ \ \ \mathrm{with} \ \ \  t_{ij}=t_{i}(T,X,C)+j\mathrm{d}t
\end{equation}

\begin{figure*}
 \vspace{1.5cm}
   \includegraphics{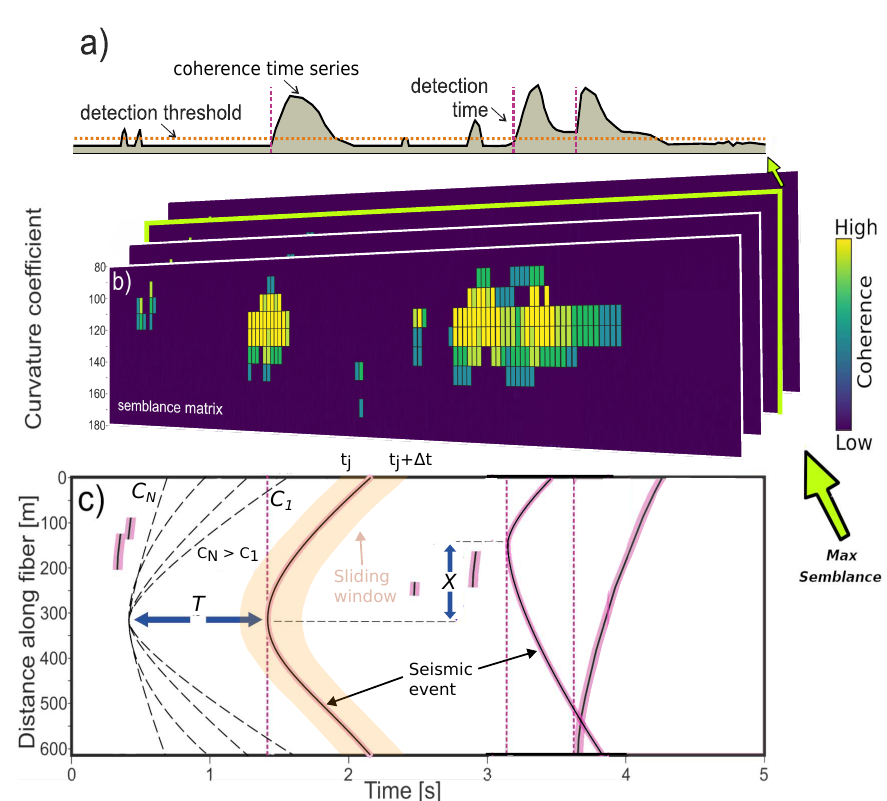}
   \caption{Sketch of the semblance-based detection method. a) represents the coherence time series, which overcomes the detection threshold in the presence of energy signal. b) shows a volume of 2D semblance matrices resulting from the scanning of the waveform coherence along geometrical hyperbolic shapes. Each semblance panel results from the scanning at each specific X position along the spatial offset. c) illustrates the range of geometrical hyperbolas (coefficients $C_1$ to $C_N$) at each $X$ and $T$ used to measure the waveform coherence along a finite-width data window ($t_j$ to $t_j+\Delta t$) (from now on sliding window) illustrated by the orange hyperbolic area. The coefficients $X$ and $T$ allow for the scanning of events along the spatial and temporal axes, respectively. In the absence of an energy signal, the method returns low coherence values. However, in the presence of a hyperbolic event, it is detected as a high coherence region, which, if overcomes the detection threshold and satisfy our clustering criteria, is declared as a seismic event.}
   \label{figure2}
\end{figure*}

Where $M$ is the total number of seismic traces (i.e. the number of the recording elements of the fiber), $N$ is the length of the sample window, and $A$ is the amplitude for the $i^{th}$ trace at the time $t_{ij}$. Depending on the application, we can choose to use either raw or processed waveforms (e.g. envelope) to calculate the semblance. The Semblance value ranges between 0 and 1, where 1 indicates perfect coherence and 0 no coherence of the signals from the different traces. The detector identifies seismic events by evaluating the waveform coherence along geometrical hyperbolic windows (sliding window hereafter, Fig. \ref{figure2}.c) while changing their curvature and the position of the vertex over the data space.

The algorithm generates a volume of 2D semblance matrices (Fig. \ref{figure2}.b) whose number of matrices and the rows and columns of each matrix is given by the number of $X$, $C$ and $T$ coefficients, respectively. The latter is decided upon the required resolution and processing times. Finally, the selected semblance matrix is the one with the highest coherence values.

The hyperbolic trajectories that do not follow the seismic wavefield will have an overall semblance value close to zero since it is equivalent to the sum of random noise. On the contrary, the hyperbolic trajectories that do follow the seismic wavefield, regardless the wave phase, will tend to have higher coherence values. The hyperbolic trajectories used to measure the waveform coherence are completely independent from external information (i.e. velocity model and hypocentral location), therefore, our method uses the waveform data as the only input. The output is a time series of coherence values that is the result of squaring and summing the coherence values along the columns of the selected semblance matrix. Squaring the values of the semblance matrix allows us to reduce even more the small coherence values related to the random noise while increasing the difference with higher coherence values, potentially associated with seismic events. Each column represents a time step in the computation of the semblance matrix and the number of columns depends on the number of $T$ coefficients (Fig. \ref{figure2}.a). The coherence time series is a 1D expression of the semblance matrix, obtained in the first part of the detection methodology. This outcome is used as input for the second part of the detection methodology, where we perform a clustering of the coherence values that exceed a detection threshold to identify potential detections of microseismic events.

It is important to note that applying the Semblance function directly to raw seismic waveforms (i.e. zero mean traces) can lead, in some cases, to spurious results. This is due to the non-isotropic radiation pattern of an earthquake source. In such cases, waveforms recorded along the fiber may have reversed polarities, thus reducing the overall coherence produced by the stacking process. To avoid this issue one may use normalized STA/LTA functions based on energy (a non-negative function), that mitigates the effect of the radiation pattern \citep{grigoli2013automated}. However, stacking non-negative functions (such as the energy or the envelope of each trace) reduces the radiation pattern's impact but does not suppress the random noise. Random noise suppression can be achieved by zero-mean stacking functions , such as raw waveforms. In the FORGE case, the length of the fiber is limited compared to the distance from the stimulation site. Such a condition allows the successful use of raw waveforms without further data processing and without having the problem of the radiation pattern effect (due to the limited sampling of the focal sphere).

\subsubsection{Clustering Analysis}

\begin{figure*}
 \vspace{1.5cm}
   \includegraphics{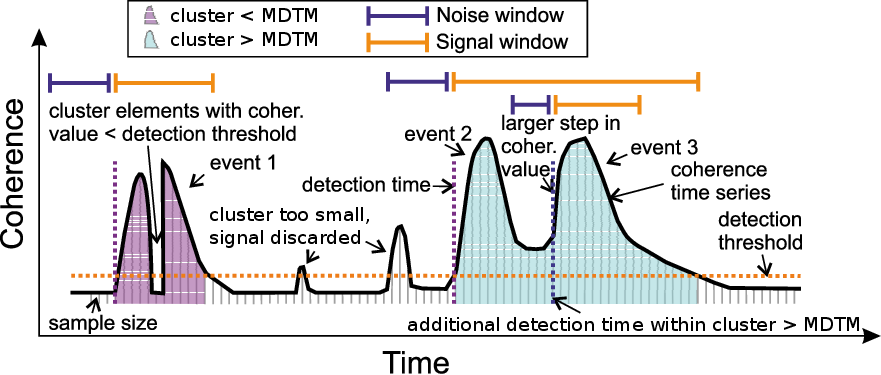}
   \caption{Schematic representation of the clustering approach used to individuate seismic events from the coherence time series (Fig. \ref{figure2}). A seismic event is returned when a cluster contains a minimum number of samples above the detection threshold and passes a SNR control (see text). The number of samples within the coherence time series is equal to the number of columns of the semblance matrix. The value of each sample of the coherence time series is equal to the sum along the columns of the coherence matrix of the squared coherence values (Fig. \ref{figure2}). When an event is detected, the detection time of the event corresponds to the time of the first sample in the coherence time series. When the cluster size is larger than a time longer than the maximum duration of the targeted microseismicity (MDTM), we inspect the possibility that a given cluster contains two events (see sub-section Clustering Analysis).}
   \label{cluster}
\end{figure*}

We calculate the detection threshold as the mean of the coherence time series after removing the 5\% of the lowest and largest values. The clustering approach requires the calibration of two parameters: 1) the minimum number of samples required to form a cluster, and 2) the maximum number of adjacent samples below the detection threshold to consider a single cluster. Clusters that meet the previous parameters are considered as potential microseismic events where the detection time corresponds to the time of the first sample in the cluster (Fig. \ref{cluster}).

To refine the initial list of detections from the cluster analysis, we apply a signal-to-noise-ratio (SNR) criterion and calculate the RMS of the signal and noise windows as follows:

\begin{equation}
E_{RMS} = \sqrt{\frac{\sum_{i=1}^{N} w_{i}^{2}}{N}}
\end{equation}

Where $N$ is the length of the time window in samples and $w_{i}$ is the $i$-th sample of the waveform. Then, the SNR is computed as:

\begin{equation}
 SNR = 20 log10 \frac{E_{RMS{signal}}}{E_{RMS{noise}}}
\end{equation}

The SNR threshold used to refine the initial list of detections will depend on the data quality and noise levels and can be manually tuned to enhance the number of real detections and minimize the number of false detections.

The signal and the noise windows are represented by a certain number of coherence samples after and before a detection time, respectively. The noise window is shorter than the signal window. We retain the detection as an event if the SNR is larger than a certain threshold, otherwise we reject it. In the case of consecutive coherence samples exceeding the detection threshold for a time longer than the maximum duration expected for the targeted microseismicity, we inspect the possibility of the presence of two microseismic events within the same cluster of coherence samples. Multiple microseismic events falling within the same cluster are expected when the inter-event distances are very short (Fig. \ref{cluster}). In the case of clusters of coherence samples exceeding the duration of targeted microseismic events, we search for the largest positive step (i.e. the largest difference in subsequent coherence values) in the coherence values and consider it as a potential microseismic event. To confirm the event, we apply a SNR criterion like the one applied for the first detection. The method is currently unable to detect more than two events should they fall within the same cluster of the coherence time series.

It is worth mentioning that this methodology is developed specifically to detect microseismic events monitored in straight fibers where the seismic wavefield can be approximated as hyperbolic events. However, HECTOR also allows to detect events originating in the far field if low curvature parameters are provided to match plane waves and a proper tuning of the clustering parameters.

In the next sections, we test and validate the performance of our detector (HECTOR) by applying it both to synthetic and real data.

\subsection{Data pre-processing}

\begin{figure*}
 \vspace{1.5cm}
 \includegraphics{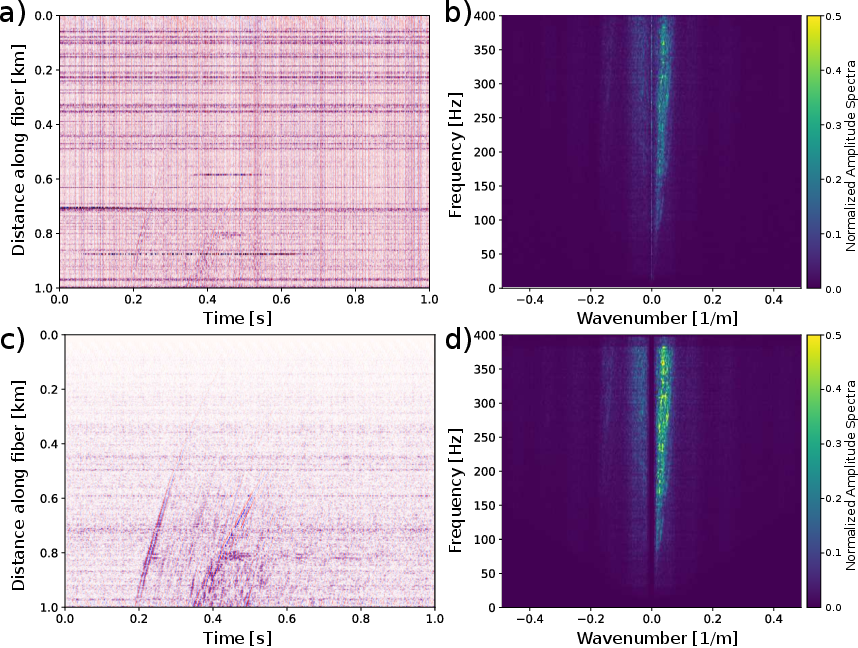}
   \caption{Example of the application of the denoising workflow. a) Raw DAS data with strong coherent noise parallel to the fiber-optic cable. b) FK spectrum of the raw DAS data. The strong coherent noise is mapped along the wavenumber zero and hinders the spectrum of the energy signal. c) Denoised DAS data. d) FK spectrum of the denoised DAS data. The color scale of the FK spectra are normalized to 1 but saturated to 0.5 for visualization purposes.}
   \label{figure3}
\end{figure*}

A sequence of pre-processing steps are applied to the data before running the detector. In this work we defined an automatic denoising workflow (Fig. \ref{figure3}) that is crucial to improve the capability of our detector especially for weak or very low-magnitude events.

The denoising workflow consists of the following steps:
1) Application of a low-pass filter (before downsampling to avoid aliasing effects) and downsampling of the traces, the latter to enhance computational speed.
2) Removal of the linear and mean trends of each trace of the array. 3) Normalization of the trace amplitudes by its corresponding maximum to reduce the effect of the geometrical spreading, receiver coupling and nonlinear effects (see \citeauthor{miah2017}., \citeyear{miah2017} for a detailed review). 
4) Application of a band-pass filter to remove the high frequency and low frequency noise on the traces and isolate the frequency band of interest for the events in the catalogue. 
5) Application of a Frequency-Wavenumber (FK) filter to attenuate any coherent noise. The FORGE dataset is characterized by strong coherent noise parallel to the fiber. The FK filter is an optimal array seismology technique used to attenuate this kind of unwanted energy.

Fig. \ref{figure3} shows an example of a low-amplitude microseismic event (maximum amplitude 44 $n\epsilon/s$) before and after applying the denoising workflow. The event is slightly visible in the time domain and in the FK spectrum of the raw data (Fig. \ref{figure3} a and b). However, after applying the denoising workflow, the coherent noise along the $k=0$ wavenumber is strongly attenuated as well as the high frequency random noise, highlighting the spatial coherence of the seismic event (Fig. \ref{figure3} c and d).

\subsection{Data}

We first generated a synthetic DAS dataset of 36 events sampled at 500 Hz, with an inter-time distance of 5 seconds, central frequency of 50 Hz and local Magnitudes (Ml) ranging between -2.0 to 1.5 by using the software Salvus \citep{salvus} and the velocity model available in \cite{lellouch2020comparison}. The synthetic dataset resembles real DAS data recorded at FORGE, and allows us to have full control over the earthquake source parameters, test the performance of HECTOR and validate our methodology.

\begin{figure*}
 \vspace{1.5cm}
 \includegraphics{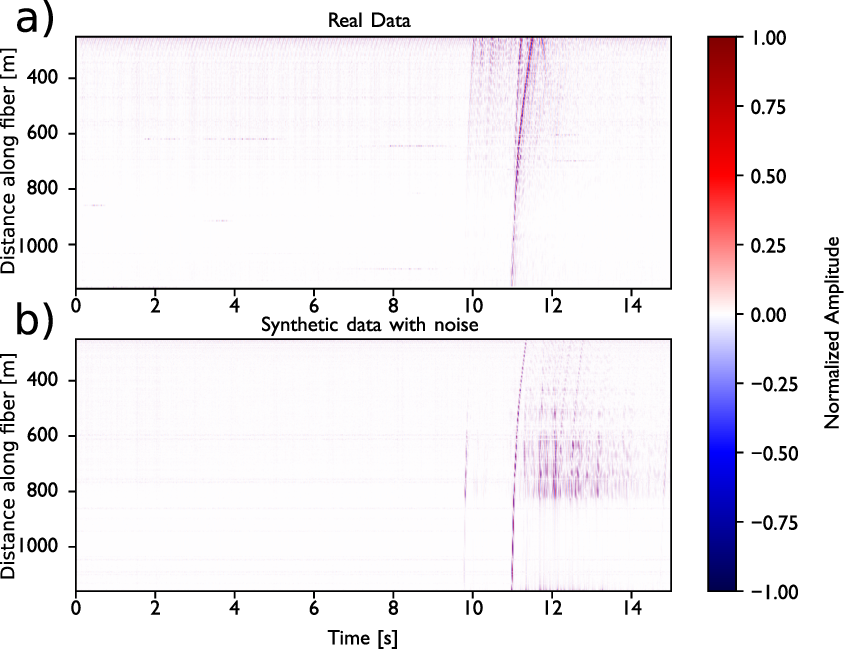}
   \caption{a) 15 seconds of real DAS data containing the Ml 0.91 earthquake occurred on April 23$^{th}$, 2019 at 21:32:09 UTC, located at a hypocentral distance of 3390 m from the very end of the fiber, used to build the synthetic waveforms. b) 15 seconds of synthetic DAS data. The synthetic event has an amplitude equivalent to a Ml 0.91 event and it is located at a depth of 1600 m and epicentral distance of 3000 m from the well.}
   \label{figure5}
\end{figure*}

To build the dataset we first generated the synthetic waveforms (Fig. \ref{figure5}) for a reference event recorded during the stimulation campaign in April 2019. Using the earthquake catalogue available in \cite{lellouch2020comparison}, we defined as reference an event of Ml 0.91 that occurred on April 23$^{th}$, 2019 at 21:32:09 UTC, at hypocentral distance of 3390 m from the bottom of the fiber. We simulated the synthetic strain waveforms for a fiber cable of 1.2 km with 1 m inter-channel spacing (as in the real dataset). We then converted the waveforms in strain-rate and corrected the amplitudes by scaling them with the waveform amplitudes of the corresponding real event. The scaling of the amplitudes is necessary to achieve synthetics that are closer to the recorded signals. 

Finally, to simulate the full set of events with different magnitudes we re-scaled the amplitudes of the synthetic waveforms using the following relation:

\begin{equation}
A_{i}^{M}(t) =  A_{i}^{R}(t) \frac{10^{M}}{10^{M^{R}}}
\label{eq7}
\end{equation}

where $A_{i}^{R}(t)$ are the waveform amplitudes of the reference event with magnitude $M^{R}$ (here $M^{R}$ = 0.91), while $A_{i}^{M}(t)$ are the scaled amplitudes for an event of magnitude $M$. The index $i$ refers to the $i$th DAS sensor. Finally, to obtain a dataset as similar as possible to the real data, we superimposed Gaussian distributed noise to the synthetic traces. We first calculate the real noise as the mean of the RMS of four 15 seconds chunks of data (i.e. four files). We then calculate the standard deviation of the data in the four 15 seconds chunks to simulate Gaussian distributed noise. Finally, we apply to the synthetic traces the denoising workflow of section 2.2 to measure the maximum amplitudes before calculating the noise. This synthetic dataset aims to reproduce different SNR conditions rather than reproducing the exact amplitudes of events of different magnitudes which is beyond the scopes of this study. 

We then evaluated the performance of HECTOR by applying it to real data acquired at FORGE during the stimulation campaign in April 2022. We used data acquired with the Silixa’s Carina sensing system along two simultaneously interrogated observation wells, namely 78A-32, and 78B-32. As mentioned before, the interrogated optical-fiber had a length of 2482 m and DAS data were acquired using a gauge length of 10 m and a channel spacing of 1 m. Data were saved with a sampling frequency of 4 kHz and in files of 15 second chunks. Additional details on the DAS acquisition can be found in the Stimulation Silixa Microseismic Report \citep{silixa}.

We focus on data acquired during the third stimulation stage, starting on April 21\textsuperscript{st} 2022 at 13:48:22 UTC, that generated the largest number of microseismic events among the three stimulation stages performed in April 2022. We downloaded 24 hours of DAS data (1.3 TB), from April 21\textsuperscript{st} 2022 at 13:00 UTC to April 22\textsuperscript{st} 2022 at 13:00 UTC. Silixa’s catalogue contains 1199 events within the investigated time interval \citep{silixa}.

\section{Results}

In this section we describe the parameters chosen for each test and report the detection results for the synthetic and real datasets, respectively.

\subsection{Results for the synthetic dataset}

A required step in the development of novel seismological algorithms consists in extensive testing with synthetic data.

Here, for our synthetic tests, we apply HECTOR with the following parameters: sliding windows ($N$) of 20 samples and steps ($T$) of 10, and hyperbolic curvature values ($C_1$ to $C_N$) between 50 and 200 with steps of 5. The curvature parameters are dependent on the distance of the events from the fiber. The size and steps of the sliding window are decided upon the sampling rate of the data and the required resolution.
We fixed the lateral search parameter ($X$) to be the depth of the deepest sensor which is a good assumption when the microseismic events are deeper than the bottom of the fiber. We set to eight the minimum number of samples over the detection threshold to declare an event and to three the maximum number of consecutive samples below the detection threshold to identify the synthetic microseismic events. The previous parameters were defined based on a trial and error procedure.

\begin{figure*}
 \vspace{1.5cm}
 \includegraphics{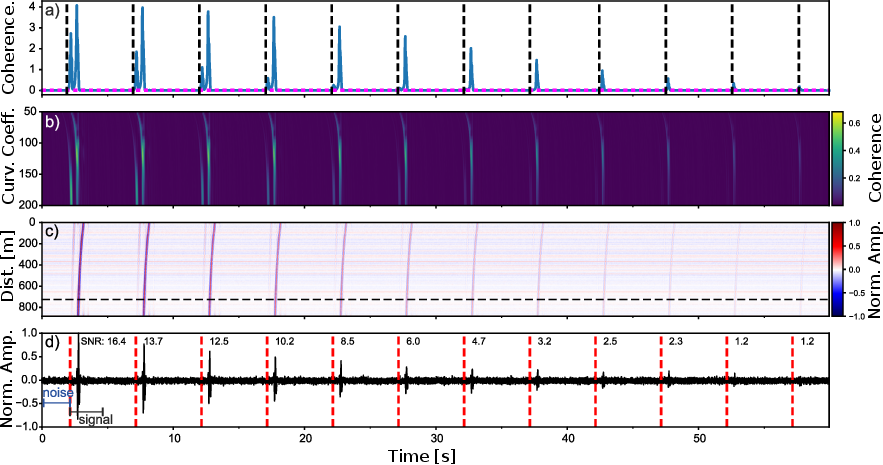}
   \caption{60 seconds of synthetic DAS data 
   containing events from Ml -0.8 to 0.3 (from right to left) with overlapped Gaussian noise. a) Coherence time series where the starting time of each event is indicated by the black vertical dashed-line. b) Semblance matrix resulting from the scanning of the waveform coherence along geometrical hyperbolic shapes. c) Synthetic event waveforms separated by 5 second windows. The dashed black line indicates the channel shown in panel d. d) Waveform from a single channel that contains the events from Ml -0.8 to 0.3 (from right to left). Top-right of the event waveforms is reported the SNR calculated on the shown trace. SNR is estimated using a noise window of 1 second before the P-wave arrival and a signal window of 1.5 seconds after the P-wave arrival.}
   \label{figure6}
\end{figure*}

We report the results of the synthetic test in Fig. \ref{figure6} and Fig. S1 of the Supplementary material. The application to the synthetic dataset clearly shows that HECTOR can successfully detect microseismic events with very low SNR or even microseismic events hidden in the noise if the signal is coherent across multiple channels of the fiber (Fig. \ref{figure6}). For the synthetic data we also observe a good agreement between the P-wave arrival of the events (Fig. \ref{figure6}c) and the detection time (Fig. \ref{figure6}a). Microseismic events with lower amplitudes (Fig. \ref{figure6}cd) result in lower coherence values (Fig. \ref{figure6}ab). However, we are unable to detect signals with extremely low amplitudes with respect to the noise amplitude as shown in Fig. S1 of the supplement. After successfully validating HECTOR with the synthetic dataset we apply it to real data.

\subsection{Results for the real dataset}

\begin{figure*}
 \vspace{1.5cm}
 \includegraphics{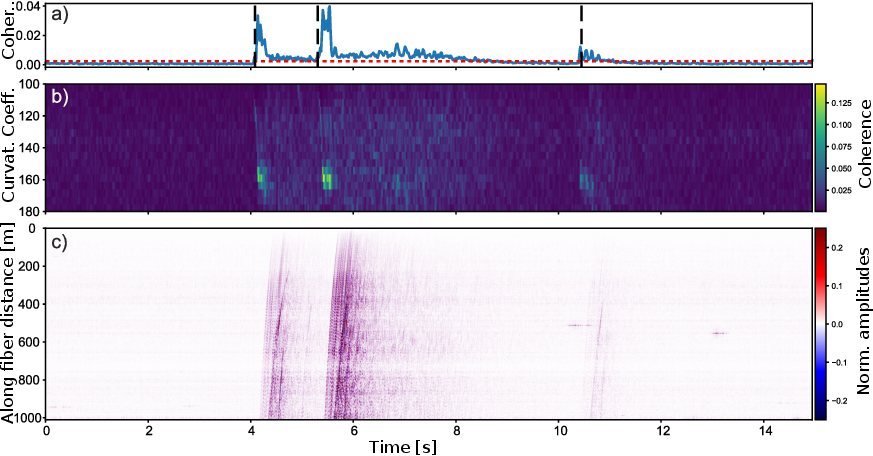}
   \caption{ a) Coherence time series for 3 events in a 15 seconds window. Detection times are: 17:06:28.5, 17:06:29.9 and 17:06:34.8 UTC on April 21\textsuperscript{st}, 2022. Dashed vertical black lines indicate the detection time of the microseismic events. b) Semblance matrix in which the three regions of maxima correspond to the three microseismic events detected by HECTOR c) Real DAS data from FORGE (well 78B-32) that contains three microseismic events with varying amplitudes. The first two events occur between 4 and 6 seconds while the third event, with the smallest amplitude among the three, occur between 10 and 11 seconds. DAS data in panel $c$ is already filtered following the denoising workflow (section 2.2).}
   \label{figure7}
\end{figure*}

\begin{figure*}
 \vspace{1.5cm}
 \includegraphics{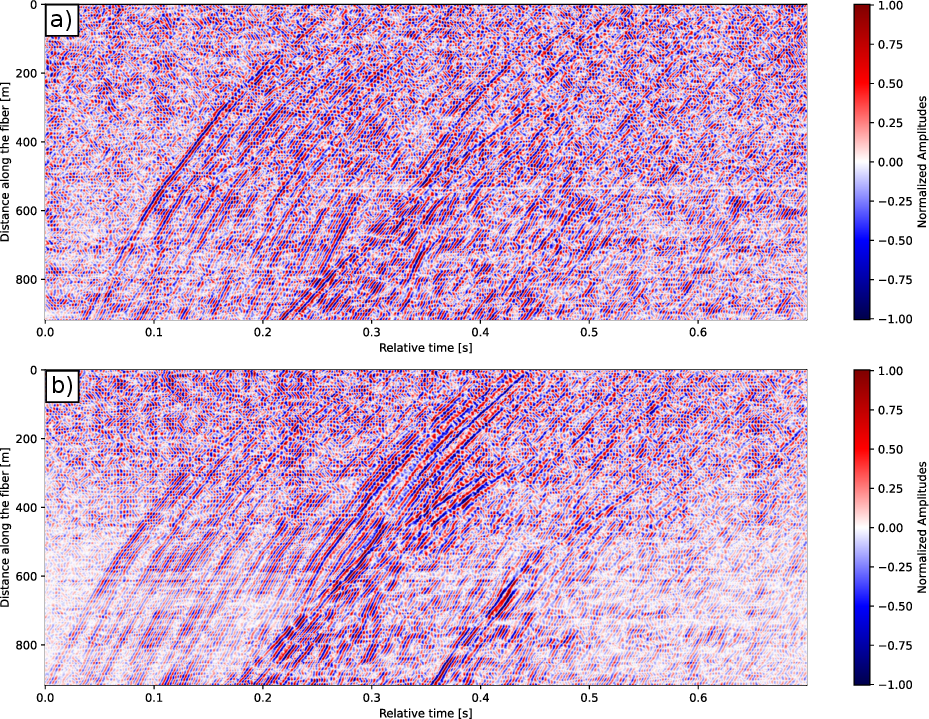}
   \caption{Example of two microseismic events present in the catalog from this study (detected by HECTOR) and not in Silixa’s catalog. a) Event with detection time on April 21$^{st}$ 2022, at 14:46:11.3 UTC. b) Event with detection time on April 21$^{st}$ 2022, at 17:47:38.3 UTC. These events have not been identified by the STA/LTA algorithm used by Silixa.}
   \label{sup1.eps}
\end{figure*}

\begin{figure*}
 \vspace{1.5cm}
 \includegraphics{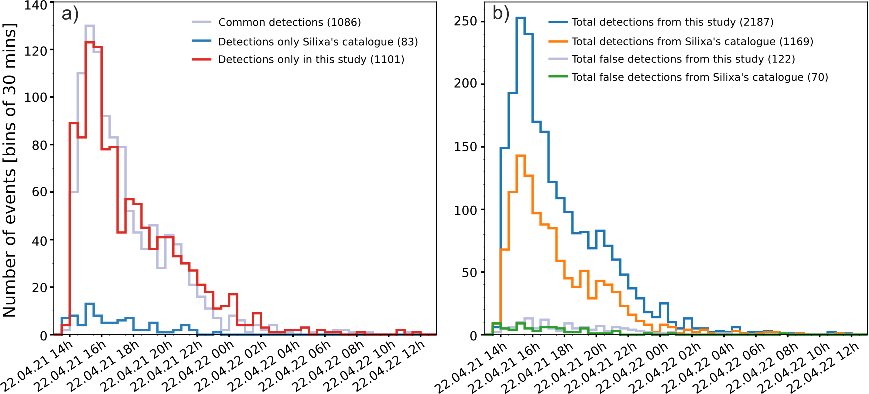}
   \caption{Detections from HECTOR applied to 24h of real data from FORGE and comparisons with an existing catalogue. Histogram bins are of 30 minutes. a) Common detections between this study and the Silixa's catalogue (gray), detections obtained only from this study (red), and the detections present only in the Silixa's catalogue (blue). b) Total detections from this study (blue), total detections in the Silixa's catalogue (orange), total false detections from this study (gray), and total false detections from Silixa's catalogue (green).}
   \label{figure8}
\end{figure*}

In order to apply the detector to the real data, we calibrated the detection and clustering parameters on a set of 50 files containing: single events, multiple events with small inter-distances (Fig. \ref{figure7}), or only noise (Fig. S2).  We observe that the typical duration of microseismic events recorded along the fiber-optic cable is 0.6-0.8 seconds. To enhance computational speed, we used only the channel interval from 1350 to 2384 (Fig. \ref{figure7}), corresponding to the optical-fiber cable located along the observational well 78B-32. We omit the topmost channels from the well 78B-32 (from 1215 to 1350) due to the high noise levels near the surface. We downsampled the real data to 600 Hz, applied a band-pass filter between 10 to 250 Hz and a FK filter to attenuate coherent linear noise and unwanted energy. 

The calibrated detection parameters for the evaluation of the waveform coherence consist of a sliding window of 20 samples with a time search step ($T$) of 10 samples and the curvature values ($C_1$ to $C_N$) from 100 to 180 with a step of 5. The lateral search was fixed to match the depth of the deepest sensor as in the case of the synthetics. This assumption is valid in the specific case of FORGE where the events are deeper than the bottom of the fiber.

Regarding the clustering parameters for the detection of microseismic events, we set to 10 the minimum number of samples above the detection threshold to identify a cluster (i.e. potential detection). If two or more consecutive clusters (or samples above the threshold) are separated by less than two samples below the detection threshold, they are grouped into the same cluster. As mentioned in the methodology, we apply a SNR criterion to refine the list of detections. We defined a signal window to be 30 samples (corresponding to 0.5 seconds) of the coherence time series and the noise window to be 20 samples (from 2 to 22 coherence samples before the detection time, corresponding to about 0.3 s) (Fig. \ref{cluster}). The second pair of signal and noise windows to explore the presence of two events within the same cluster are equal to the first 20 samples of the coherence time series after the additional detection time, while the noise window includes 15 samples, namely from 2 to 17 samples before the additional detection time, Fig. \ref{cluster}). We kept the detection as an event if the SNR was larger than 4.

After the calibration, we processed the 24 hours of real data and initially detected 2236 events (before de-clustering). As comparison, Silixa’s catalogue contains 1199 detections for the same time interval (before de-clustering). Their catalogue contains microseismic events detected using an STA/LTA method applied to the fiber-optic cables deployed in both the observation wells.

To better show the performance of our algorithm, we made a rigorous comparison of the events in our catalogue with those in the Silixa’s catalogue \citep{silixa}. We removed possible duplicated events by de-clustering both catalogues with the removal of the second detection when two detections are less or equal than 0.7 seconds apart. We retained 2187 and 1169 events, respectively. Then, we searched for common events in the catalogues by considering a time interval of \textpm 0.6 s from the detection time of the events in our catalogue. We found 1086 common events, while 83 events are only contained in the Silixa’s catalogue, and 1101 events are only contained in our catalogue (Fig. \ref{sup1.eps}). We manually inspected all the detected events and classified as "false detection" 6 out of the 1086 events common to both catalogues, 64 out of the 83 events only present in the Silixa’s catalogue, and 116 of the 1101 events only present in our catalogue. We classified as "false detection" the coda of microseismic events, events with not clearly visible P and/or S arrivals, false detections, and distant seismic events (Fig. S3).

In summary, our catalogue contains 2187 microseismic events (Fig. \ref{figure8}), from which 122 are classified as "false detection" (5.6 \%), while the Silixa’s catalogue contains 1169 microseismic events with 70 of them classified as "false detection" (6.0 \%). The reported total number of events also includes common detections to both catalogues. The pre-processing workflow and the semblance-based detection method proposed in this study enable the detection of a number of events that is significantly larger (almost double) than those detected with a standard STA/LTA method.

\section{Discussion and Conclusions}

In this work we developed HECTOR, a microseismic event detection method based on the analysis of the spatial coherence of seismic wavefields along geometrical hyperbolic trajectories.

After defining a proper DAS denoising workflow to improve the detection rate, we first validated our detection method with synthetic DAS waveforms that resemble the real data of the FORGE geothermal site recorded during the April 2019 campaign, and established a detection rate of microseismic events close to a SNR of 0.1 (Figs. \ref{figure5} and S4). We then applied the detector to 24 hours of real data acquired during the third stimulation stage at FORGE during the April 2022 acquisition campaign. With our method, we detected a total of 2187 microseismic events, from which only 5.6 \% are false detections (2065 real detections).

We have demonstrated that our detection method outpaces traditional pick-based detection methods like STA/LTA for microseismic monitoring with DAS data, as we almost doubled the number of detected events using the same dataset. During the same period, Geo Energie Suisse, using borehole geophones deployed in well 58-32, closer to the stimulation well (905 m) with respect to well 78B-32 (1308 m) used in this study (Fig. \ref{figure4}), derived a catalogue of 1431 well-locatable events with magnitudes ranging from 0.52 to -1.8 \citep{oedi_1399}. It is worth mentioning that Geo Energie Suisse reported a preliminary total number of about 18400 detections during the stimulation stage 3 \citep{oedi_1429}. A direct comparison with the seismic catalogues obtained with borehole geophones, as we did for the catalogue obtained from DAS data (\citep{silixa}), is challenging and beyond the scope of this paper because of the different locations of the monitoring wells.

In this study we provide a methodology that exploits the complexity of the seismic wavefield in such a detail only possible to be recorded with DAS, to perform an accurate detection of microseismicity, often characterized by a massive number of small, noise contaminated, seismic events with short inter-event times and without the need of external information (i.e. velocity models). Our approach supports the potential benefits of turning any production or stimulation well into an observation well too by just installing an optical-fiber behind the casing, which would significantly reduce operational costs.

Our detection algorithm has been developed specifically to detect microseismic events monitored in straight fibers, where the seismic wavefield can be approximated as hyperbolic events. However, HECTOR also allows to detect events originating in the far field if low curvature parameters are provided to match plane waves and a proper tuning of the clustering parameters.

The parameter $X$ allows for the scanning of microseismic events along hyperbolic trajectories at different positions along the spatial offset (parallel to the fiber-optic cable), even beyond the distance covered by the DAS system. In the cases where an $X$ position, which represents the vertex of the hyperbola, is beyond the fiber-optic cable, the hyperbolic trajectory used to measure the spatial coherence will be only a part of the hyperbola (i.e. the part covered by the fiber-optic cable). For each $X$ value, a semblance matrix ($C-T$) is computed. Hence, we compute $N$ semblance matrices, where $N$ represents the number of $X$ values considered in the lateral search parameter. Among these $C-T$ semblance matrices, it is then selected the one having the highest maximum coherence value. In this way the detector can be applied to approximately straight fibers independently on the hypocentral position of the event. However, lateral search also implies calculating a number of $C-T$ semblance matrices equal to all the possible vertex positions along the spatial offset, resulting in a higher computational cost. This limitation will be improved by introducing parallelization in a future version of the algorithm. Nonetheless, with the parameters defined in this study, the algorithm requires a computational time of approximately 0.25 seconds per each second of data, measured in a laptop with an Intel quad-core i7 processor and 16 GB of Random Access Memory (RAM).

Fig. S4 illustrates the effect of the parameter $X$ for three synthetic events at different hypocentral locations with respect to a vertical well of 1 km depth. We see that in all the cases, the events are successfully detected. The event in Fig. S4c highlights how the detector successfully identifies both P and S arrivals. Besides them, it also shows at least another detection related to a best-fitting hyperbola having a lower curvature coefficient. This further detection can be explained by considering that the algorithm tests several potential hyperbolas having different curvatures. For a homogeneous and isotropic half-space having constant velocity, the signal at the fiber would be a perfect hyperbola. However, when dealing with 1D velocity models, the signal recorded at the fiber can be seen as a combination of different hyperbolas with different curvatures. As a result, for the same event and seismic phase, there may be more than one detection related to hyperbolas having different curvatures and fitting different portions of the signal. Hence, multiple detections associated with the same event can be filtered out by imposing a minimum interval between consecutive detections. 
As we observe in Figure S5a,b, sometimes the detection is a few hundred milliseconds anticipated. There may be two main reasons for this time difference. The first is due to the best-fitting hyperbola's vertex position along the time axis. Given a borehole installation and a source beneath the fiber (as in Fig. S5b), the best-fitting hyperbola vertex is located at a depth D greater than the fiber depth and at a time instance that precedes the arrival time of the signal at the fiber. The detection time corresponds to the hyperbola vertex position along the time-axis. Hence, the time difference between the vertex time position and the signal arrival time at the fiber is the observed delay. On the other hand, in Figure S5a, the anticipated detection is due to the clustering algorithm. The detector identifies clusters of consecutive coherence samples whose amplitude exceeds a threshold. If a cluster is classified as an event, the detection is set at the cluster starting time, which may not correspond to the maximum of the coherence times series for that cluster.

Despite the aforementioned limitations, we have demonstrated that our waveform-based and clustering detection method outpaces traditional pick-based detection methods and makes of HECTOR a suitable methodology for DAS-based real-time microseismicity monitoring in industrial operations.

\begin{acknowledgments}
We would like to thank the FORGE project for making the DAS data used in this study freely available \citep{oedi_1379}. This work is part of a collaboration between the Department of Earth Sciences of the University of Pisa and the Swiss Seismological Service in the framework of the De-Risking Enhanced Geothermal Energy Projects.(Innovation for DEEPs). DEEP is subsidized through the Cofund GEOTHERMICA, which is supported by the European Union’s HORIZON 2020 programme for research, technological development, and demonstration under Grant Agreement Number 731117. This work is also supported by the University of Pisa under the Progetto di Ricerca di Ateneo PRA Fluid migration in the upper crust: from natural hazards to geo-resources (project number 274 PRA 2022 66).
G.M.B was funded by the Volkswagen Foundation and by the RDSME of the Ruhr University of Bochum.
\end{acknowledgments}

\begin{dataavailability}
The list of detections from this study and the comparison with the Silixa's detections for the same period is freely available in \citep{bocchini_gian_maria_2023_5106469}.
The detector will be made publicly available on GitHub upon acceptance of the paper.
\end{dataavailability}


\bibliographystyle{gji}
\bibliography{references} 

\end{document}